\documentclass[structabstract]{aa}  
\usepackage{graphicx}
\usepackage{txfonts}
\usepackage{natbib}
\bibpunct{(}{)}{;}{a}{}{,} 

\begin{document}
   \title{Effects of rotation on the evolution and asteroseismic properties of red giants}

   \author{P. Eggenberger\inst{1,2} \and A. Miglio\inst{1,}\thanks{Postdoctoral Researcher, Fonds de la Recherche Scientifique - FNRS, Belgium} \and J. Montalban\inst{1} \and O. Moreira\inst{1} \and A. Noels\inst{1} \and G. Meynet\inst{2} \and A. Maeder\inst{2}}

   \institute{Institut d'Astrophysique et de G\'eophysique de l'Universit\'e de Li\`ege, 17 All\'ee du 6 Ao\^ut, B-4000 Li\`ege, Belgique\\
              \email{[eggenberger;miglio;montalban;moreira;noels]@astro.ulg.ac.be}
\and
Observatoire de Gen\`eve, Universit\'e de Gen\`eve, 51 Ch. des Maillettes, CH-1290 Sauverny, Suisse\\ 
\email{[georges.meynet;andre.maeder]@unige.ch}
             }

   \date{Received; accepted}

 
  \abstract
   {The recent observations of solar-like oscillations in many red giant stars with the CoRoT satellite stimulated the theoretical study of the effects of various physical processes on the modelling of these stars.}
   {The influence of rotation on the properties of red giants is studied in the context of the asteroseismic modelling of these stars.}
   {The effects of rotation on the global and asteroseismic properties of red giant stars with a mass larger than the mass limit 
    for degenerate He burning are investigated by comparing rotating models computed with a comprehensive treatment 
    of shellular rotation to non-rotating ones.}
   {While red giants exhibit low surface rotational velocities, we find that the rotational history of the star has a large impact on its properties 
   during the red giant phase. In particular, for stars massive enough to ignite He burning in non-degenerate conditions, 
   rotational mixing induces a significant increase of the stellar luminosity and shifts the location of the core 
   helium burning phase to a higher luminosity in the HR diagram. This of course results in a change of the seismic properties 
   of red giants at the same evolutionary state. As a consequence the inclusion of rotation significantly changes 
   the fundamental parameters of a red giant star as determined by performing an asteroseismic calibration. 
   In particular rotation decreases the derived stellar mass and increases the age. Depending on the rotation law assumed 
   in the convective envelope and on the initial velocity of the star, 
   non-negligible values of rotational splitting can be reached, which may complicate the observation and identification 
   of non-radial oscillation modes for red giants exhibiting moderate surface rotational velocities.
   By comparing the effects of rotation and overshooting, we find that the main-sequence widening and the increase 
   of the H-burning lifetime induced by rotation ($V_{\rm ini}=150$\,km\,s$^{-1}$) are well reproduced by non-rotating models with an overshooting parameter 
   of 0.1, while the increase of luminosity during the post-main sequence evolution is better reproduced 
   by non-rotating models with overshooting parameters twice as large. This illustrates the fact that rotation not only increases 
   the size of the convective core but also changes the chemical composition of the radiative zone.}
   {}

   \keywords{stars: red giants -- stars: rotation -- stars: oscillations -- stars: fundamental parameters}

   \maketitle
%

\section{Introduction}

The solar five-minute oscillations have provided a wealth of information
on the internal structure of the Sun. These results stimulated various attempts to detect
a similar signal on other stars. In past years, solar-like oscillations have been observed 
for a handful of stars either from the ground by using stabilized spectrographs developed 
for extra-solar planet searches or from photometric data obtained from space \citep[see e.g.][]{bed07}. 
While these asteroseismic observations mainly focused on solar-type stars, solar-like oscillations 
have also been detected for a few red giants \citep{fra02, bar04, bar07, der06}. These first detections together 
with the recent clear identification of non-radial oscillations for many red giant stars (G--K giants) with CoRoT \citep{der09} 
stimulated the theoretical study of the asteroseismic properties of red giant stars and of 
the effects of various physical processes on the modelling of these stars. Rotation is one of the key processes 
that changes all outputs of stellar models with a peculiarly strong impact on the physics and evolution of massive stars \citep[see e.g.][]{mae00b}. 
In this paper, we focus on the effects of rotation on the asteroseismic modelling of red giants massive enough to ignite He burning in non-degenerate
conditions by comparing stellar models including shellular rotation (i.e. a strongly anisotropic turbulence
leads to an essentially constant angular velocity on the isobars, see Sect.~\ref{phys_rot} for more details) to non-rotating models. The influence of rotation on 
the asteroseimic features of red giants and on the determination of the global properties of a red giant star 
obtained by performing an asteroseimic modelling is studied. 

The physical description of rotation is first briefly presented in Sect.~2. The results of the comparisons between rotating and non-rotating models of red giants are discussed in Sect.~3, while the conclusion is given in Sect.~4.

\section{Physical description of rotation}
\label{phys_rot}

In the radiative interiors of rotating stars, meridional circulation is generated as a result of the thermal
imbalance induced by the breaking of the spherical symmetry \citep{edd25, vog26}. This 
large scale circulation transports matter and angular momentum. Differential rotation then takes
place in the radiative zones, making the stellar interior highly turbulent. The turbulence is very anisotropic,
with a much stronger geostrophic-like transport in the horizontal than in the vertical direction 
\citep{zah92}.
The horizontal turbulent coupling favours an essentially constant angular velocity $\Omega$ on the isobars.
With this hypothesis of shellular rotation, every quantity depends solely
on pressure and can be split into a mean value and its latitudinal perturbation
\begin{equation}
f(P,\theta) = \overline{f}(P) + \tilde{f}(P)P_2(\cos \theta) \, ,
\label{shellular}
\end{equation}
where $P_2(\cos \theta)$ is the second Legendre polynomial.
In this context, the transport of angular momentum obeys an advection-diffusion equation written in Lagrangian
coordinates \citep{zah92, mae98}:
\begin{equation}
  \rho \frac{{\rm d}}{{\rm d}t} \left( r^{2}\Omega \right)_{M_r} 
  =  \frac{1}{5r^{2}}\frac{\partial }{\partial r} \left(\rho r^{4}\Omega
  U(r)\right)
  + \frac{1}{r^{2}}\frac{\partial }{\partial r}\left(\rho D r^{4}
  \frac{\partial \Omega}{\partial r} \right) \, , 
\label{transmom}
\end{equation}
$r$ being a characteristic radius, $\rho$ the mean density on an isobar and $\Omega(r)$ the mean angular velocity at level $r$.
The vertical component $u(r,\theta)$ of the velocity of the meridional circulation at a distance
$r$ to the center and at a colatitude $\theta$ can be written
\begin{equation}
u(r,\theta)=U(r)P_2(\cos \theta)\,.
\end{equation}
Only the radial term $U(r)$ appears in Eq. (\ref{transmom});
its expression is given by \citep{mae98}
\begin{eqnarray}
U(r)  =  \frac{P}{\rho g C_{P} T [\nabla_{\rm ad}-\nabla + (\varphi/\delta)
  \nabla_{\mu}]}
 \cdot  \left\{ \frac{L}{M}(E_{\Omega }+E_{\mu}) \right\}\,. 
\label{vmer}
\end{eqnarray}
$P$ is the pressure, $C_P$ the specific heat, $E_{\Omega}$ and $E_{\mu}$ are terms
depending on the $\Omega$- and $\mu$-distributions respectively, up to the third
order derivatives and on various thermodynamic quantities \cite[see][ for more details]{mae98}. 

Meridional circulation and shear mixing are considered as the main mixing mechanisms in radiative zones.
The first term on the right-hand side of Eq. (\ref{transmom}) describes the advection of angular momentum by 
meridional circulation, while the second term accounts for the transport of angular momentum by shear instability
with $D=D_{\rm shear}$. The expression of this diffusion coefficient is given by
\begin{eqnarray}
D_{\rm shear} & = & \frac{ 4(K + D_{\mathrm{h}})}
{\left[\frac{\varphi}{\delta} 
\nabla_{\mu}(1+\frac{K}{D_{\mathrm{h}}})+ (\nabla_{\mathrm{ad}}
-\nabla_{\mathrm{rad}}) \right] } \nonumber\\
&  & \times \frac{H_{\mathrm{p}}}{g \delta} \; 
 \frac{\alpha}{4}\left( 0.8836 \, \Omega{{\rm d}\ln \Omega \over {\rm d}\ln r} \right)^2 \, ,
\label{dshear}
\end{eqnarray}
with $K$ the thermal diffusivity \citep{mae01}. 
$D_{\rm h}$ is the diffusion coefficient associated
to horizontal turbulence. The usual expression for this coefficient is, according to \cite{zah92},
\begin{equation}
D_{\rm h} = \frac{1}{c_{\rm h}} r |2V(r)-\alpha U(r)|\,,
\label{Dhzahn}
\end{equation}
where $U(r)$ is the vertical component of the meridional circulation velocity, $V(r)$ the
horizontal component, $c_{\rm h}$ a constant of order unity and 
$\alpha=\frac{1}{2} \frac{{\rm d} \ln r^2 {\bar \Omega}}{{\rm d} \ln r}$. 
The full solution of Eq. (\ref{transmom}) taking into account $U(r)$ and $D$ gives the non-stationary solution of the problem. 
We recall here that meridional circulation is treated as a truly advective
process in the Geneva evolution code.

The vertical transport of chemicals
through the combined action of vertical advection and strong
horizontal diffusion 
can be described as a pure diffusive process \citep{cha92}.
The advective transport is thus replaced by a diffusive term, 
with an effective
diffusion coefficient
\begin{equation}
D_{\rm eff} = \frac{|rU(r)|^2}{30D_{\rm h}}\,.
\label{Deff}
\end{equation} 
The vertical transport of chemical elements then
obeys a diffusion equation which, in addition to this macroscopic transport,
also accounts for (vertical) turbulent transport:
\begin{eqnarray}
\rho  \frac{{\rm d} c_i}{{\rm d} t} = \dot{c_i}
 + \frac{1}{r^2}\frac{\partial}{\partial r}\left[ r^2\rho
   (D_{\rm eff}+D_{\rm shear})\frac{\partial c_i}{\partial r} \right]\,,
\label{transchem}
\end{eqnarray}
where $c_i$ is the mean concentration of element $i$ and $\dot{c_i}$ represents the variations of chemical
composition due to nuclear reactions.

\section{Results}

The stellar evolution code used for these computations is the Geneva code
that includes a comprehensive treatment of shellular rotation as briefly described in the preceding section \citep[see][ for more details]{egg08}. 

\subsection{Effects of rotation in the HR diagram}
\label{sec_dhr}

To investigate the effects of rotation on the properties of red giants, we first compute the evolution of 
a 3\,M$_{\odot}$ star with and without rotation. These models are computed with a solar chemical composition as given by \cite{gre93}, 
a solar calibrated value for the mixing-length parameter ($\alpha_{\odot}=1.59$ with the input physics
used for these computations) and without overshooting from the convective core into the surrounding radiatively 
stable layers. Both models share therefore exactly the same initial parameters except for the inclusion of shellular rotation. The rotating model  
is computed with an initial velocity of 150\,km\,s$^{-1}$ on the zero age main sequence (ZAMS).

\begin{figure}[htb!]
\resizebox{\hsize}{!}{\includegraphics{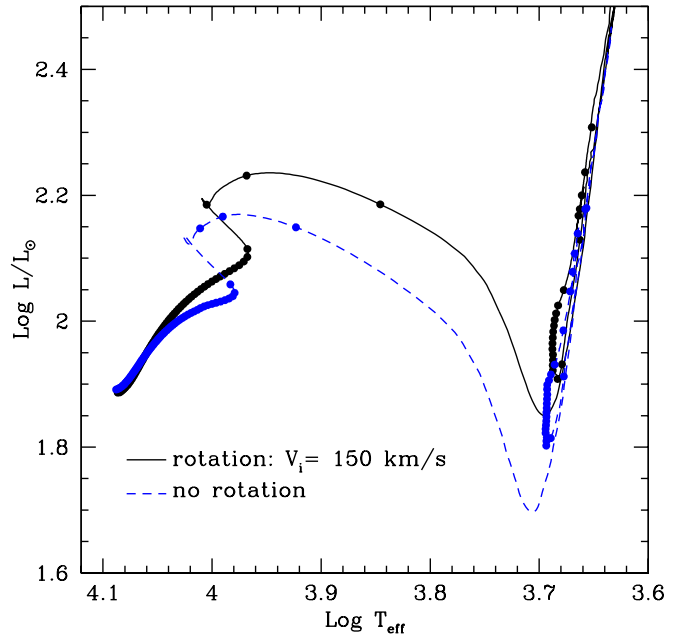}}
 \caption{Evolutionary tracks in the HR diagram for 3\,M$_\odot$ models. The continuous and dashed lines correspond to models computed with and without rotation, respectively.  Dots indicate time intervals of 5\,Myr.}
  \label{dhr_rot}
\end{figure}

Figure~\ref{dhr_rot} shows the evolutionary tracks in the HR diagram for both models. During the red giant phase, 
the star spends most of its lifetime in the loop corresponding to the core helium-burning phase. 
Consequently, it is much more likely to observe a red giant star during the helium-burning phase 
than during its rapid evolution on the red giant branch in the hydrogen shell-burning phase. 
The inclusion of rotation is found to significantly change the evolutionary tracks in the HR diagram. 
Rotating models indeed exhibit larger luminosities than non-rotating ones. In the red giant branch, the core helium-burning phase is thus 
located at a larger luminosity when rotation is included.

\begin{figure}[htb!]
\resizebox{\hsize}{!}{\includegraphics{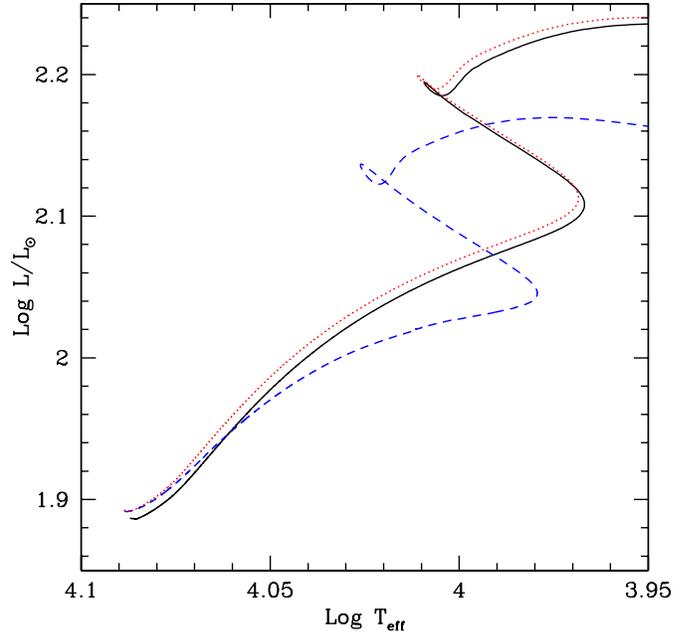}}
 \caption{Zoom on the main-sequence evolution corresponding to the evolutionary tracks shown in Fig.~\ref{dhr_rot}. The dotted line shows the evolutionary track for a model of 3\,M$_\odot$ including only rotational mixing.}
  \label{dhr_cor}
\end{figure}

In order to investigate whether these differences are mainly due to rotational mixing or to hydrostatic corrections due to rotation, 
another rotating model of 3\,M$_\odot$ with an initial velocity of 150\,km\,s$^{-1}$ on the ZAMS is computed by including only the effects of rotational mixing. 
Figure~\ref{dhr_cor} compares the main-sequence evolution of this model with the ones shown in Fig.~\ref{dhr_rot}. 
The evolutionary track in the HR diagram corresponding to the rotating model including only rotational mixing is very similar 
to the track of the model computed with a full treatment of rotation. The change of the evolutionary tracks in the HR diagram 
induced by rotation for red giants is thus mainly due to rotational mixing with only a very limited contribution from the effects of the centrifugal force.
This is due to the fact that the kinetic rotational energy of the star is much lower than its gravitational energy 
(typical ratio of 0.1\% in the present case).
 
Comparing the dotted and continuous lines in Fig.~\ref{dhr_cor}, we see that the inclusion of the hydrostatic effects of rotation 
results in a slight shift of the track towards lower luminosities and effective temperatures. This is caused by
the effective gravity of the star being reduced by the inclusion of the centrifugal acceleration term. The evolutionary track of
a star including the hydrostatic effects of rotation is thus similar to the one of a star computed with a slightly lower initial mass and
without these effects, hence the shift to the red part of the HR diagram. By comparing the non-rotating track (dashed line) in Fig.~\ref{dhr_cor} 
to the one including the full treatment of shellular rotation (continuous line), we note that only hydrostatic effects 
of rotation are observed at the beginning of the main sequence. As the evolution proceeds, rotational mixing begins 
to play a dominant role though by bringing fresh hydrogen fuel into the convective core and by transporting helium and other H-burning products 
into the radiative zone. This mixing results in a widening of the main sequence with a significant increase of the stellar luminosity for rotating models compared to non-rotating ones.

\subsection{Rotational velocities}
\label{sec_rot}

During the main-sequence evolution, the surface rotational velocity slowly decreases. The 3\,M$_\odot$ model computed with an initial velocity 
of 150\,km\,s$^{-1}$ on the ZAMS exhibits a surface velocity of 115\,km\,s$^{-1}$ at the end of the main sequence. Then the surface velocity rapidly decreases 
during the quick evolution from the blue to the red part of the HR diagram. At the bottom of the red giant branch, the velocity equals 13\,km\,s$^{-1}$ 
and reaches a mean value of about 6\,km\,s$^{-1}$ during the evolution as a red giant star. For a model computed with a lower initial velocity 
of 50\,km\,s$^{-1}$, a surface velocity of 4.7\,km\,s$^{-1}$ is reached at the bottom of the red giant branch with a mean value of about 2\,km\,s$^{-1}$ 
during the red giant phase. It is worthwhile to note here that these values are obtained by assuming a solid body rotation in the convective zones. 
While this assumption seems reasonable during the main-sequence evolution as indicated by the solar case, it becomes questionable during further 
evolution in the red giant phase where the star is characterised by an extended convective envelope \citep[see e.g.][]{swe79, den00, pal06, pal07}.  

Following the work of \cite{swe79} and \cite{pal06}, another limiting case for the rotation law in the convective envelope is taken into account 
in addition to the assumption of solid body rotation: the assumption of uniform specific angular momentum. A model of a 3\,M$_\odot$ red giant star 
with an initial velocity of  150\,km\,s$^{-1}$ is then computed by assuming solid body rotation in the convective envelope during the main sequence 
and uniform specific angular momentum after the main sequence. For this model, the assumption of uniform specific angular momentum 
in the convective envelope after the main sequence leads to a decrease of the surface velocities. At the bottom of the red giant branch, 
the velocity equals 1\,km\,s$^{-1}$ and reaches a mean value of about 0.3\,km\,s$^{-1}$ during the core helium phase on the red giant branch. 
We see that for both limiting assumptions on the rotation law in the convective envelope, a stellar model computed with a significant 
initial rotational velocity will have a low surface velocity during the red giant phase in good agreement with the observed rotational 
velocities of red giants \citep[e.g.][]{deM99, mas08}. 
We consequently conclude that while a red giant 
exhibits a low surface rotational velocity, it may have been rotating much more rapidly during the main sequence. 
Since rotation significantly changes the stellar properties during the main sequence, the further evolution in the red giant phase 
is found to significantly depend on the rotational history of the star.

\begin{figure}[htb!]
\resizebox{\hsize}{!}{\includegraphics{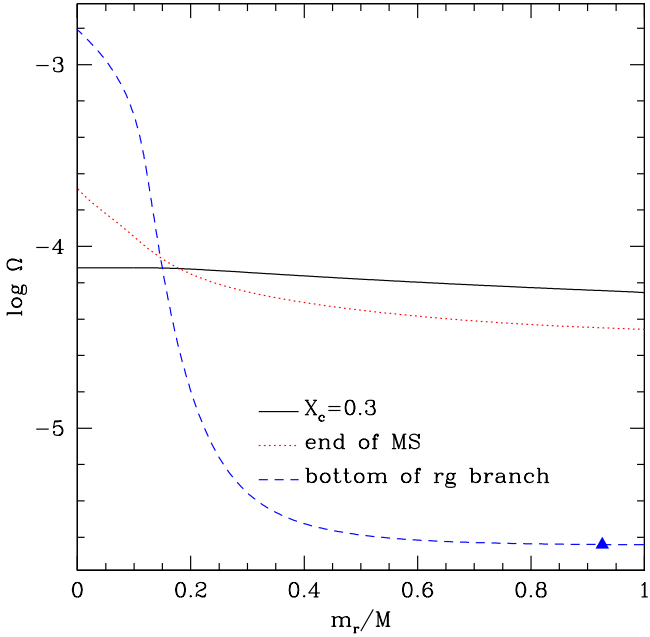}}
 \caption{Evolution of the internal rotation profile in a 3\,M$_\odot$ star with an initial velocity of 150\,km\,s$^{-1}$. 
X$_{\rm c}$ is the hydrogen mass fraction at the center. The triangle indicates the border of the convective envelope for the
model at the bottom of the red giant branch. A constant angular velocity is assumed in the convective zone.}
  \label{rot}
\end{figure}

\begin{figure}[htb!]
\resizebox{\hsize}{!}{\includegraphics{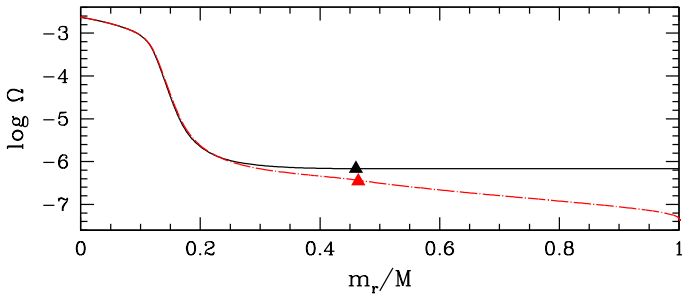}}
 \caption{Internal rotation profile for a 3\,M$_\odot$ red giant with a luminosity of 100\,L$_{\odot}$. Continuous and dashed-dotted lines correspond to a model computed with the assumption of solid body rotation and conservation of specific angular momentum in the convective envelope after the main-sequence phase, respectively. Both models are computed with an initial velocity of 150\,km\,s$^{-1}$. 
Triangles indicate the border of the convective envelope.}
  \label{rot_EC}
\end{figure}

Figure~\ref{rot} shows the evolution of the internal rotation profile for the 3\,M$_{\odot}$ model with an initial velocity of 150\,km\,s$^{-1}$ 
computed with the assumption of solid body rotation in the convective envelope. During the main sequence, a small degree of differential rotation 
is found as a result of the transport mechanisms at work in stellar interiors. Near the end of the main sequence, 
the angular velocity in the central layers grows rapidly, since the central contraction dominates the evolution of the angular momentum. 
After the main sequence and during the red giant phase, the evolution of the internal rotation profile is mainly dominated by 
the local conservation of angular momentum. This is due to the fact that secular transport mechanisms have little time to operate. 

Figure~\ref{rot_EC} compares the internal rotation profile of red giant models of 3\,M$_{\odot}$ with a luminosity of 100\,L$_{\odot}$ 
computed with two different assumptions regarding the rotation law in the convective envelope.  
When a constant angular velocity is assumed in the convective zone, the rotation profile is approximately flat between the surface and about 0.2\,$M$, 
where $M$ is the total mass of the star. 
With the assumption of a uniform specific angular momentum in the convective envelope, a slight monotonic decrease of the angular velocity 
is observed in the same region when the distance to the stellar centre increases.

\subsection{Lifetimes}

\begin{figure}[htb!]
\resizebox{\hsize}{!}{\includegraphics{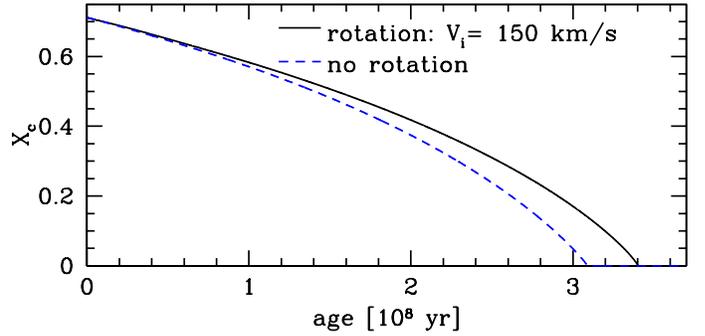}}
 \caption{Hydrogen mass fraction at the center of the star (X$_{\rm c}$) as a function of the age. The continuous line corresponds to a rotating model of 3\,M$_\odot$ with an initial velocity of 150\,km\,s$^{-1}$, while the dashed line corresponds to a model computed without rotation.}
  \label{xct}
\end{figure}

In addition to changing the evolutionary tracks in the HR diagram, rotation has also a large impact 
on the age of a given stellar model. As briefly mentioned in Sect.~\ref{sec_dhr}, rotational mixing is indeed 
found to bring fresh hydrogen fuel to the central stellar core. Figure~\ref{xct} shows the evolution of 
the central mass fraction of hydrogen (X$_{\rm c}$) as a function of the age for a 3\,M$_\odot$ star 
computed with and without rotation. At a given age, the central hydrogen mass fraction 
is larger for the rotating model than for the model without rotation. 

As a result of rotational mixing, the main-sequence lifetime is larger for stellar models including rotation. 
For instance, a 3\,M$_\odot$ star with an initial velocity of 150\,km\,s$^{-1}$ exhibits an age of 0.34\,Gyr 
at the end of the main sequence, while the corresponding non-rotating model has a lower age of 0.31\,Gyr.

\subsection{Rotation and overshooting}

\begin{figure}[htb!]
\resizebox{\hsize}{!}{\includegraphics{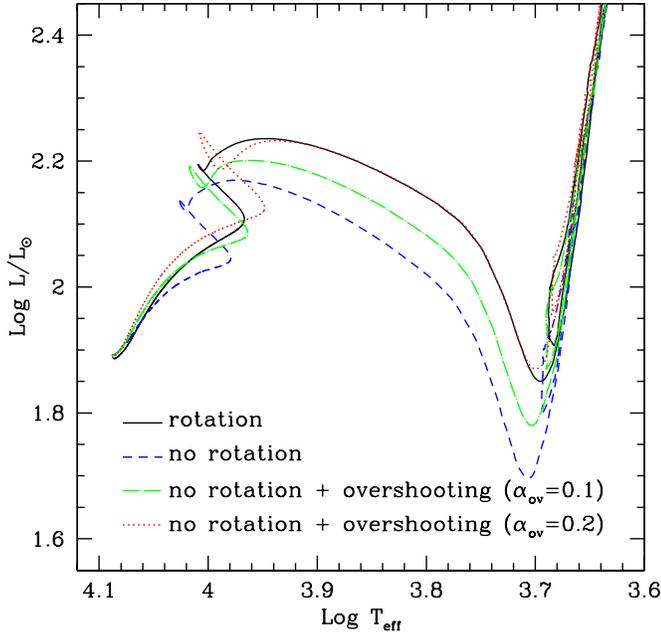}}
 \caption{Evolutionary tracks in the HR diagram for 3\,M$_\odot$ models. The continuous and dashed lines indicate models with an initial velocity of 150\,km\,s$^{-1}$ and without rotation, respectively. 
Both models are computed without overshooting. The dashed-dotted and the dotted lines correspond to non-rotating models with an overshoot parameter of 0.1 and 0.2, respectively.}
  \label{dhr_ov}
\end{figure}

\begin{figure}[htb!]
\resizebox{\hsize}{!}{\includegraphics{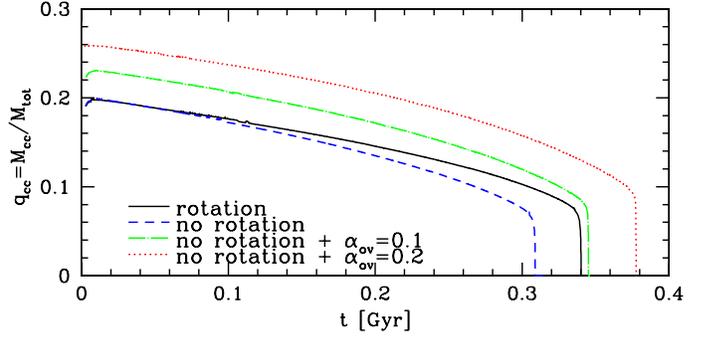}}
 \caption{Ratio of the mass of the convective core to the total mass of the star ($q_{\mathrm{cc}}$) as a function of the age for models of 3\,M$_{\odot}$. The continuous and dashed lines indicate models with an initial velocity of 150\,km\,s$^{-1}$ and without rotation, respectively. Both models are computed without overshooting. The dashed-dotted and the dotted lines correspond to non-rotating models with an overshoot parameter of 0.1 and 0.2, respectively.}
  \label{qcct}
\end{figure}

The effects of rotation and overshooting from the convective core into the surroundings layers on a distance $d_{\mathrm{ov}} \equiv \alpha_{\mathrm{ov}} \min[H_p,r_{\mathrm{core}}]$ \citep{mae89}
are compared by computing the evolution of non-rotating models of 3\,M$_{\odot}$ with an overshoot parameter $\alpha_{\mathrm{ov}}$ of 0.1 and 0.2. 
The evolutionary tracks of these models in the HR diagram are shown in Fig.~\ref{dhr_ov}. The inclusion of overshooting results in an increase of the luminosity of the star. 
The turn-off point at the end of the main-sequence occurs at a lower effective temperature and a higher luminosity, thereby increasing the main-sequence width. 
These effects are due to the increase of the mass of the convective core when overshooting is included in the computation as shown in Fig.~\ref{qcct}. 
As a result of the larger convective cores, the main sequence lifetime is increased by about 10\% and 20\% for models with an overshoot parameter of 0.1 and 0.2, respectively. 

\begin{figure}[htb!]
\resizebox{\hsize}{!}{\includegraphics{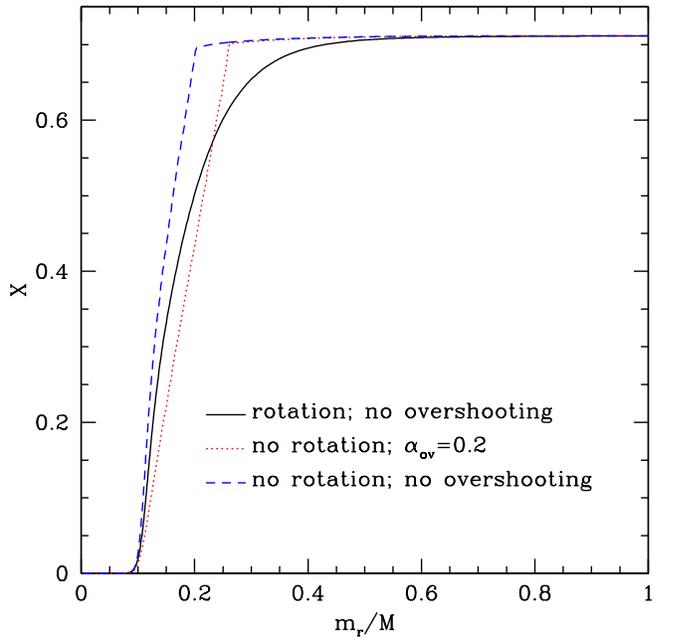}}
 \caption{Hydrogen abundance profile for 3\,M$_{\odot}$ models with the same effective temperature ($\log(T_{\rm eff})=3.9$) during the post-main sequence phase of evolution.}
  \label{profx}
\end{figure}

As discussed in Sect.~\ref{sec_dhr}, rotation is also found to increase the luminosity of the star and to extend the main sequence tracks towards 
lower effective temperatures. This is essentially due to two effects. First, rotational mixing brings fresh hydrogen fuel into the convective core, 
slowing down its decrease in mass during the evolution on the main-sequence. Secondly, rotational mixing 
transports helium and other H-burning products into the radiative zone.  
The first effect can be clearly seen in Fig.~\ref{qcct} and results in a shift of the turn-off point at the end of the main-sequence 
to lower effective temperatures. The second effect is illustrated in Fig.~\ref{profx}. 
As shown in Figs.~\ref{dhr_ov} and \ref{qcct}, the main-sequence widening as well as the increase 
of the H-burning lifetime due to rotational mixing is well reproduced by a non-rotating model with an overshoot parameter $\alpha_{\rm ov}=0.1$.
Fig.~\ref{dhr_ov} also clearly shows that the increase of the luminosity is larger for the rotating model than for the non-rotating model 
with an overshoot parameter of  $\alpha_{\rm ov}=0.1$. In particular, the luminosity during the post-main sequence 
phase of evolution of the rotating model is better reproduced by a non-rotating model with a larger overshoot parameter of 0.2. 
This is due to the multiple effects of rotational mixing (i.e. an increase of the size of the convective core and a change of the
chemical composition in the radiative zone) that lead to a significant increase of the stellar luminosity. Indeed, Fig.~\ref{profx} shows that the chemical composition of
the internal layers of the radiative zone is changed when rotational effects are included. In particular, the transport of helium 
into the radiative zone leads to a larger mean molecular weight and hence to a larger luminosity for the rotating model.

We thus conclude that the inclusion of overshooting 
has a similar effect on the global stellar parameters as rotation. 
In particular, both processes lead to an increase of the stellar luminosity and age together with an extension of the main-sequence to lower effective temperatures. 
It is however interesting to note that all the changes of the stellar global properties induced by rotation cannot be reproduced 
by a single non-rotating model with a given value of the overshooting parameter. Indeed, we find that the increase of the age and the main-sequence widening 
due to rotation is approximately reproduced by a non-rotating model with a moderate overshooting $\alpha_{\rm ov}=0.1$, 
while the increase of the stellar luminosity induced by rotational mixing (in particular during the post-main sequence phase of evolution) 
is better reproduced by a non-rotating model with a larger value of the overshooting parameter $\alpha_{\rm ov}=0.2$.

\subsection{Effects of rotation on asteroseismic properties of red giants}

\begin{figure}[htb!]
\resizebox{\hsize}{!}{\includegraphics{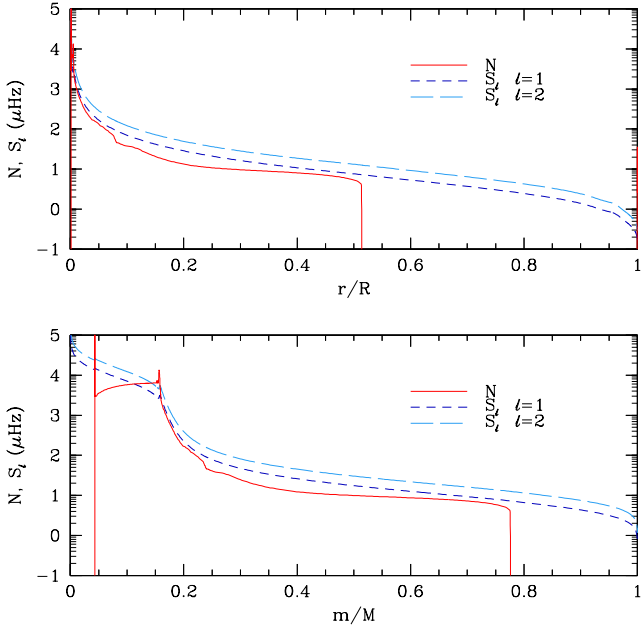}}
 \caption{Characteristic frequencies of a 3\,M$_{\odot}$ star 
in the helium burning phase. $N$ and $S_{\ell}$ correspond to the Brunt-V\"ais\"al\"a and the
Lamb frequencies, respectively.}
  \label{propa}
\end{figure}

Red giant stars in the helium burning phase are characterised by a deep convective envelope 
with a very small convective core resulting from helium burning in the center.
The large density near the centre of the star results in very large values 
of the Brunt-V\"ais\"al\"a frequency in the central layers. This is illustrated in Fig.~\ref{propa} which
shows the characteristic frequencies in the interior of a 3\,M$_{\odot}$ star 
in the helium burning phase. $N$ and $S_{\ell}$ correspond to the Brunt-V\"ais\"al\"a and the
Lamb frequencies, respectively. We recall here that the asymptotic analysis of the oscillation equations 
shows that an oscillating solution is found when one of the two following conditions is fullfilled:
the angular mode frequency is larger than $N$ and $S_{\ell}$, or its frequency is lower 
than both $N$ and $S_{\ell}$ \cite[e.g.][]{osa75}. The first condition corresponds to 
acoustic or p modes, while the second one corresponds to gravity or g modes.

\begin{figure}[htb!]
\resizebox{\hsize}{!}{\includegraphics{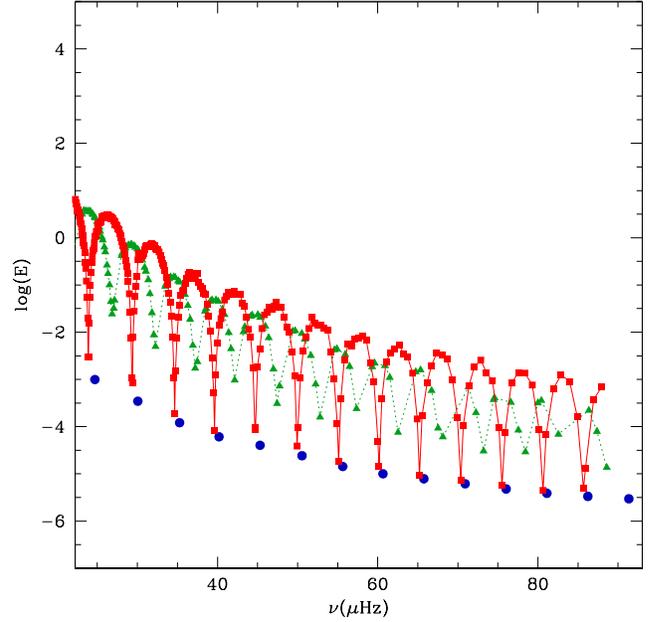}}
 \caption{Mode inertias for the 3\,M$_{\odot}$ model in the helium burning phase shown in Fig.~\ref{propa}. 
Dots, triangles and squares correspond to modes with $\ell=0$, $\ell=1$ and $\ell=2$, respectively.}
  \label{inertie}
\end{figure}

Due to the large increase of the Brunt-V\"ais\"al\"a frequency in the central layers,
the situation is somewhat complex in the case of a red giant star. Figure~\ref{propa} indeed shows that
a mode can simultaneously exhibit a g-mode-like behaviour in the core (its frequency being lower than
$N$ and $S_{\ell}$) and p-mode-like behaviour in the envelope (frequency larger than both
$N$ and $S_{\ell}$). Oscillation modes of mixed p-mode and g-mode 
properties are thus expected in red giants. As can been seen in Fig.~\ref{inertie}, 
these non-radial modes are dominated in most cases by the g-mode behaviour, and therefore 
exhibit a high inertia. Indeed, most of the $\ell=1$ and $\ell=2$ modes
have an inertia significantly larger and hence surface amplitudes significantly smaller 
than the purely acoustic radial modes. However, for non-radial modes trapped 
in the acoustic cavity the inertia decreases and can become similar to the one of the radial modes. 
For $\ell=1$ modes, Fig.~\ref{propa} shows that the evanescent region (where the mode frequency is larger than
$N$ but lower than $S_{\ell=1}$) is narrow, so that the p- and g-mode regions are not well 
separated. Consequently, $\ell=1$ modes are not efficiently trapped in the acoustic cavity, and the inertia 
of $\ell=1$ modes remains larger than for radial modes.
For $\ell=2$ modes, the evanescent region is broader and
the separation between the p- and g-mode region is then sufficient to obtain oscillation modes, 
which in terms of inertia are very similar to purely acoustic modes. As a result, Fig.~\ref{inertie} shows that 
there is a trapped $\ell=2$ mode with a dominant amplitude close to every radial mode, defining thereby 
the small separation between $\ell=0$ and $\ell=2$ modes \cite[][]{chr04b}. 
Note that this discussion is only based on inertia consideration independently from 
the fact that non-radial modes in red giants could be more strongly damped than the radial modes \citep[see e.g.][]{dzi01}. 
This important aspect of solar-like oscillations in red giant stars is examined in detail in \cite{dup09}.

\subsubsection{Models with the same initial mass}
\label{sec_astero_Mcte}

In the precedings sections, the effects of rotation on the global properties of red giants have been discussed. 
As shown in Fig.~\ref{dhr_rot}, a red giant star spends most of its time in the loop corresponding 
to the core helium-burning phase. In order to study the effects of rotation on the asteroseismic properties of red giants, 
we thus compare rotating and non-rotating models during the helium-burning phase. 
The main effect of rotation is then to shift the location of this phase to larger luminosities. 
Let us compare the asteroseismic features of rotating and non-rotating models computed with exactly 
the same initial parameters and situated at the same evolutionary stage during the central helium burning phase 
(i.e. with the same value of the central helium mass fraction Y$_{\rm c}$).

\begin{figure}[htb!]
\resizebox{\hsize}{!}{\includegraphics{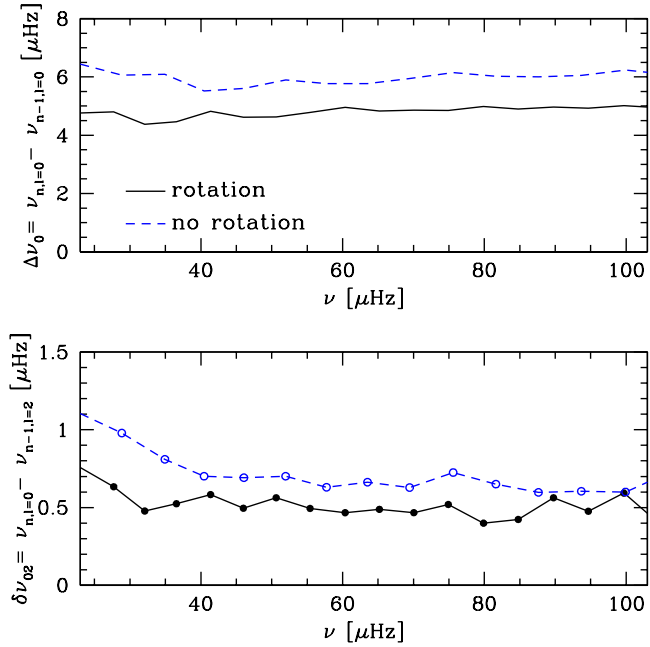}}
 \caption{Large separation for radial modes and small separation between $\ell=0$ and trapped $\ell=2$ modes versus frequency for two red giant models of 3\,M$_\odot$ at the same evolutionary stage during the helium burning phase. The continuous and dashed lines correspond to a model computed with an initial velocity of 150\,km\,s$^{-1}$ and without rotation, respectively.}
  \label{gdpt_rot}
\end{figure}

For this purpose, a red giant model of 3\,M$_\odot$ with an initial velocity of 150\,km\,s$^{-1}$ 
located in the middle of the helium burning phase (Y$_{\rm c}=0.355$) is compared to 
the corresponding non-rotating red-giant model computed with the same initial parameters 
except for the inclusion of shellular rotation. The rotating red giant model is characterised 
by an age of 0.41\,Gyr, a luminosity $L^{\rm rot}=96.9$\,L$_{\odot}$ and 
an effective temperature $T_{\rm eff}^{\rm rot}=4862$\,K. The red giant model computed without rotation 
has a lower age and luminosity of respectively 0.40\,Gyr and $L^{\rm no rot}=75.5$\,L$_{\odot}$, 
with an effective temperature $T_{\rm eff}^{\rm no rot}=4930$\,K. 
The theoretical low-$\ell$ frequencies of the models are then computed 
by using the Aarhus adiabatic pulsation code \citep{chr97}. 
The asteroseismic properties of both red giant models are compared in Fig.~\ref{gdpt_rot}. 
The values of the large separation are calculated from the radial modes, while the small separations 
between $\ell=0$ and $\ell=2$ modes are determined by considering only the $\ell=2$ modes 
with an inertia close to the radial modes which are trapped in the acoustic cavity.   

Figure~\ref{gdpt_rot} shows that the inclusion of rotation results in a clear decrease of the large separation. 
The rotating red giant model exhibits a mean large separation of 4.83\,$\mu$Hz, while the non-rotating model 
has a larger mean separation of 5.99\,$\mu$Hz. The mean large separation is mainly proportional to the square root of the star's mean density.
Both models sharing the same mass, this difference of about 20\% is due to the larger radius 
of the rotating model (13.9\,R$_{\odot}$ instead of 11.9\,R$_{\odot}$ for the non-rotating model).
Since the effective temperatures of the models computed with and without rotation are very similar (the difference is about 1\%), 
the decrease of the mean large separation with rotation can be directly related to the large increase 
of the stellar luminosity induced by rotational mixing (a difference of about 25\% between both models). 

Concerning the small separation between $\ell=0$ and $\ell=2$ modes trapped in the acoustic cavity of the star, 
a similar decrease is observed when rotation is taken into account (see bottom of Fig.~\ref{gdpt_rot}). 
For main-sequence stars, the small separation is known to be mainly sensitive to the conditions in the central regions 
of the star. However, it also retains some sensitivity to the near-surface structure. \cite{rox03} 
therefore introduce the use of another asteroseismic quantity: the ratio of the small to large separation. 
This ratio constitutes a better diagnostic of the central parts of a star than the small separation, 
since it is essentially independent of the structure of the outer layers and is determined solely 
by the interior structure \citep{rox03,rox05}. In the case of the present red giants models, 
the decrease of the values of the small separation when rotation is included is similar to the decrease 
of the large separation, so that the ratio between the small and large separation is approximately 
the same for the rotating and non-rotating model. Thus, the different values of the small separation for a red giant model 
computed with and without rotation mainly reflects the change of the global stellar properties 
and not a change in the structure of the central regions of the star.   

\begin{figure}[htb!]
\resizebox{\hsize}{!}{\includegraphics{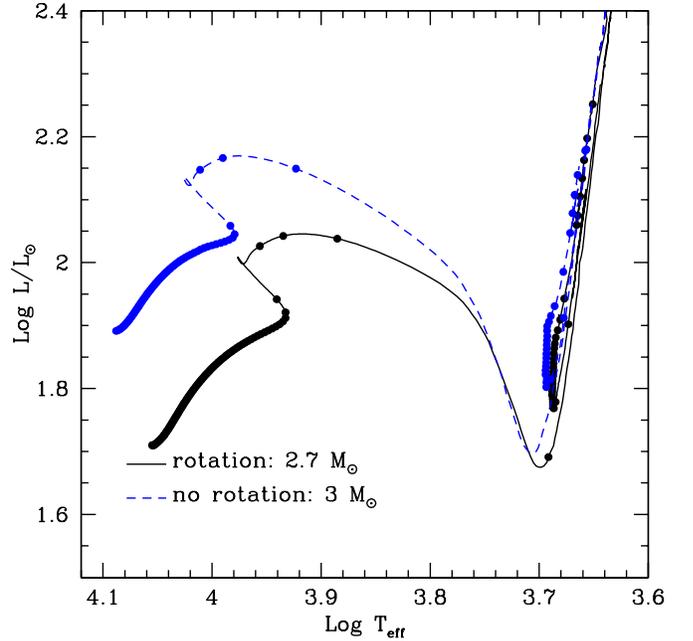}}
 \caption{Evolutionary tracks in the HR diagram for rotating and non-rotating models of red giants sharing approximately the same luminosity during the helium burning phase. The continuous line corresponds to a 2.7\,M$_{\odot}$ star computed with an initial velocity of 150\,km\,s$^{-1}$, while the dashed line corresponds to a non-rotating model of 3\,M$_{\odot}$. Dots indicate time intervals of 5 Myr.}
  \label{dhr_m2p7}
\end{figure}

Finally, the values of rotational splittings predicted for rotating models of red giants are discussed 
by comparing the two models of 3\,M$_{\odot}$ with a luminosity of 100\,L$_{\odot}$ presented 
in Sect.~\ref{sec_rot} and Fig.~\ref{rot_EC}, which are computed with two different assumptions 
regarding the rotation law in the convective envelope.
A mean rotational splitting of about 0.1\,$\mu$Hz is found for the model computed with a solid body rotation in the convective envelope 
for trapped $\ell=2$ modes, while a lower value of about 0.03\,$\mu$Hz is found in the case of uniform specific angular momentum.
This is of course due to the lower values of the internal rotational velocity in the acoustic region 
of the star for the model computed with the assumption of a uniform specific angular momentum 
in the convective envelope (see Fig.~\ref{rot_EC}).
We thus see that depending on the rotation law assumed in the convective envelope (and of course on the value of the initial velocity of the star), 
the value of the mean rotational splitting during the red giant phase can be quite different. It is interesting to note that the splitting of about 
0.1\,$\mu$Hz found for the model computed with a constant angular velocity in the convective zone is not negligible compared to the low
values predicted for the large and small separations of red giants (see for instance Fig.~\ref{gdpt_rot}). Indeed, a rotational splitting of 0.1\,$\mu$Hz
leads to a frequency spacing of 0.4\,$\mu$Hz between modes $\ell=2$ modes with $m=+2$ and $m=-2$, which is only slightly lower than the mean value of 
the small separation between $\ell=0$ and $\ell=2$ modes trapped in the acoustic cavity of the star (about 0.6\,$\mu$Hz). Due to rotational splitting, 
the observation and identification of non-radial oscillation modes regularly spaced in frequency will then be much more difficult  
in red giants exhibiting a moderate surface rotational velocity (about 6\,km\,s$^{-1}$ for the model studied here) 
than in red giants with lower surface velocities.

\subsubsection{Models with the same luminosity}

After studying the effects of rotation on asteroseismic properties of red giant stars 
by comparing rotating and non-rotating models computed with the same initial parameters, 
we are now interested in investigating the effects of rotation on the determination 
of the fundamental stellar parameters and asteroseismic properties for red giants sharing 
the same location in the HR diagram. For this purpose, another model is computed in order 
to obtain rotating red giants in the helium burning phase with approximately 
the same luminosity as the non-rotating red giant stars of 3\,M$_{\odot}$. 
As shown in Fig.~\ref{dhr_m2p7}, the red giant branch of a 2.7\,M$_{\odot}$ star 
with an initial velocity of 150\,km\,s$^{-1}$ shares the same luminosity interval 
as the red giant branch of a non-rotating star of 3\,M$_{\odot}$. 
In particular, the luminosity of the bottom of the red giant branch and of the central helium burning loop 
is approximately the same for both models. The inclusion of rotation results therefore in a decrease 
of the mass of the star (of approximately 10\%) in order to reproduce the location of the helium burning phase 
of red giants in the HR diagram. At a given luminosity, a rotating red giant model exhibits 
also a slightly lower effective temperature than a non-rotating model (with a mean difference of about 90\,K). 
This is due to the lower mass required by the rotating models to reach the same luminosity during the red giant phase.
The lower initial mass of models including rotation results also in a large increase of the determined age for a red giant. 
During the core helium burning phase, rotating models exhibit indeed an age about 40\% larger than non-rotating models. 
We thus see that rotation significantly changes the global stellar parameters required to reproduce the same location of 
the helium-burning red giant stars in the HR diagram.   

\begin{figure}[htb!]
\resizebox{\hsize}{!}{\includegraphics{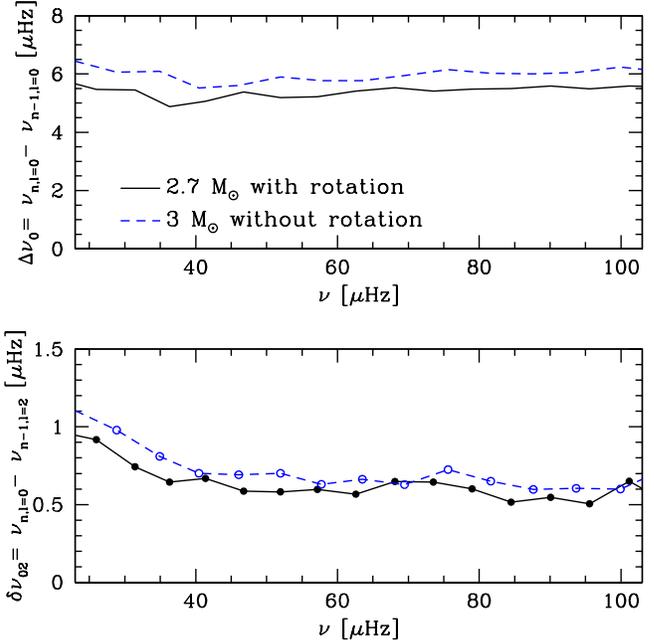}}
 \caption{Large separation for radial modes and small separation between $\ell=0$ and trapped $\ell=2$ modes versus frequency for two red giant models with the same luminosity during the helium burning phase. The continuous and dashed lines correspond to a 2.7\,M$_{\odot}$ model computed with an initial velocity of 150\,km\,s$^{-1}$ and a 3\,M$_{\odot}$ star without rotation, respectively.}
  \label{gdpt_m2p7}
\end{figure}

To investigate the effects of rotation on the asteroseismic properties of red giants with the same luminosity, 
theoretical low-$\ell$ frequencies of the rotating 2.7\,M$_{\odot}$ red giant in the helium burning phase are computed. 
It is characterised by an age of 0.578\,Gyr, a luminosity of 75.4\,L$_{\odot}$ 
and an effective temperature of 4844\,K (with Y$_{\rm c}=0.229$). 
The asteroseismic features of both models are compared in Fig.~8. The rotating 2.7\,M$_{\odot}$ model exhibits 
a lower value of the mean large separation than the non-rotating model (5.43 instead of 5.99\,$\mu$Hz). 
This difference of about 10\% mainly results from the different mean density of both models due to their different masses. 
There is also a small contribution from the slight change 
in the effective temperature of the rotating model. Both models share indeed the same luminosity, while the slightly 
lower effective temperature of the rotating model implies a larger value of the stellar radius resulting 
in a slight decrease of the mean large separation of the rotating model. 
The values of the small separation between $\ell=0$ and $\ell=2$ modes trapped in the acoustic cavity of the star 
are also found to decrease when rotation is included (see bottom of Fig. 8). As discussed above, 
this difference for red giant stars computed with and without rotation mainly reflects the change 
of the global stellar properties (particularly the mass of the star in this case) and not a change 
in the structure of the central regions of the star.

\section{Conclusion}

While red giants generally exhibit low surface rotational velocities, they may have been rotating much more rapidly during the main sequence. 
Since rotation significantly changes the stellar properties during the main sequence, 
the further evolution in the red giant phase is found to depend on the rotational history of the star. 
In particular, the comparison of rotating and non-rotating models computed with exactly the same initial parameters 
except for the inclusion of shellular rotation shows that rotation induces a significant increase of the stellar luminosity. 
The change of the evolutionary tracks in the HR diagram due to rotation is mainly due to rotational mixing with 
only a very limited contribution from hydrostatic corrections induced by rotation. 
In the case of red giant stars massive enough to ignite He burning in non-degenerate conditions, 
rotation shifts the location of the core helium burning phase to higher luminosity in the HR diagram. 
This of course results in a change of the global asteroseismic properties of red giants situated at the same evolutionary stage. 
Lower values of the large separation of radial modes as well as of the small separation between $\ell=0$ and $\ell=2$ modes trapped in the acoustic cavity 
of the star are then found for rotating models than for the non-rotating ones. 

The effects of rotation and overshooting have also been compared. We find that both processes increase the stellar luminosity and the determined age 
and shift also the location of the end of the main sequence to lower effective temperatures. However, the changes of the stellar global properties 
induced by rotation cannot be reproduced by a single non-rotating model with a given value of the overshooting parameter. 
This is due to the fact that, contrary to overshooting, the changes of the global stellar properties induced by rotation are not only due 
to the increase of the size of the convective core, but also to the transport of helium and H-burning products by shear mixing and meridional circulation in the radiative layers. 

Finally, the effects of rotation on the determination of the fundamental stellar parameters and asteroseismic properties for red giants sharing the same location in the HR diagram have been studied.
Rotation is then found to decrease the determined value of the mass of a red giant located at a given luminosity in the HR diagram (by about 10\% for a typical value of the rotation velocity) 
and to increase the value of the age (by about 40\% for a typical rotation rate). We thus conclude that although red giants are slow rotators, 
the inclusion of rotation significantly changes the fundamental parameters determined for a star from an asteroseismic calibration. 
This is especially interesting in the context of the recent observations of solar-like oscillations in red giants obtained with CoRoT \citep{der09} 
and of new asteroseismic data expected from ground based observations and upcoming space missions.

\begin{acknowledgements}
We would like to thank J. Christensen-Dalsgaard for providing us with the Aarhus adiabatic pulsation code.
P.E. is thankful to the Swiss National Science Foundation for support.
\end{acknowledgements}

\bibliographystyle{aa} 
\bibliography{biblio} 

\begin{thebibliography}{29}
\expandafter\ifx\csname natexlab\endcsname\relax\def\natexlab#1{#1}\fi

\bibitem[{{Barban} {et~al.}(2004){Barban}, {de Ridder}, {Mazumdar}, {Carrier},
  {Eggenberger}, {de Ruyter}, {Vanautgaerden}, {Bouchy}, \& {Aerts}}]{bar04}
{Barban}, C., {de Ridder}, J., {Mazumdar}, A., {et~al.} 2004, in ESA Special
  Publication, Vol. 559, SOHO 14 Helio- and Asteroseismology: Towards a Golden
  Future, ed. D.~{Danesy}, 113

\bibitem[{{Barban} {et~al.}(2007){Barban}, {Matthews}, {de Ridder}, {Baudin},
  {Kuschnig}, {Mazumdar}, {Samadi}, {Guenther}, {Moffat}, {Rucinski},
  {Sasselov}, {Walker}, \& {Weiss}}]{bar07}
{Barban}, C., {Matthews}, J.~M., {de Ridder}, J., {et~al.} 2007, \aap, 468,
  1033

\bibitem[{{Bedding} \& {Kjeldsen}(2007)}]{bed07}
{Bedding}, T.~R. \& {Kjeldsen}, H. 2007, Communications in Asteroseismology,
  150, 106

\bibitem[{{Chaboyer} \& {Zahn}(1992)}]{cha92}
{Chaboyer}, B. \& {Zahn}, J.-P. 1992, \aap, 253, 173

\bibitem[{{Christensen-Dalsgaard}(1997)}]{chr97}
{Christensen-Dalsgaard}, J. 1997, $\mathtt{http://astro.phys.au.dk/\! \sim \!
  jcd/adipack.n/}$

\bibitem[{{Christensen-Dalsgaard}(2004)}]{chr04b}
{Christensen-Dalsgaard}, J. 2004, \solphys, 220, 137

\bibitem[{{de Medeiros} \& {Mayor}(1999)}]{deM99}
{de Medeiros}, J.~R. \& {Mayor}, M. 1999, \aaps, 139, 433

\bibitem[{{De Ridder} {et~al.}(2009){De Ridder}, {Barban}, {Baudin}, {Carrier},
  {Hatzes}, {Hekker}, {Kallinger}, \& {Weiss}}]{der09}
{De Ridder}, J., {Barban}, C., {Baudin}, F., {et~al.} 2009, Nature, 459, 398

\bibitem[{{De Ridder} {et~al.}(2006){De Ridder}, {Barban}, {Carrier},
  {Mazumdar}, {Eggenberger}, {Aerts}, {Deruyter}, \& {Vanautgaerden}}]{der06}
{De Ridder}, J., {Barban}, C., {Carrier}, F., {et~al.} 2006, \aap, 448, 689

\bibitem[{{Denissenkov} \& {Tout}(2000)}]{den00}
{Denissenkov}, P.~A. \& {Tout}, C.~A. 2000, \mnras, 316, 395

\bibitem[{{Dupret} {et~al.}(2009){Dupret}, {Belkacem}, {Samadi}, {Montalban},
  {Moreira}, {Miglio}, {Godart}, {Ventura}, {Ludwig}, {Grigahc{\`e}ne},
  {Goupil}, {Noels}, \& {Caffau}}]{dup09}
{Dupret}, M., {Belkacem}, K., {Samadi}, R., {et~al.} 2009, \aap, 506, 57

\bibitem[{{Dziembowski} {et~al.}(2001){Dziembowski}, {Gough}, {Houdek}, \&
  {Sienkiewicz}}]{dzi01}
{Dziembowski}, W.~A., {Gough}, D.~O., {Houdek}, G., \& {Sienkiewicz}, R. 2001,
  \mnras, 328, 601

\bibitem[{{Eddington}(1925)}]{edd25}
{Eddington}, A.~S. 1925, The Observatory, 48, 73

\bibitem[{{Eggenberger} {et~al.}(2008){Eggenberger}, {Meynet}, {Maeder},
  {Hirschi}, {Charbonnel}, {Talon}, \& {Ekstr{\"o}m}}]{egg08}
{Eggenberger}, P., {Meynet}, G., {Maeder}, A., {et~al.} 2008, \apss, 316, 43

\bibitem[{{Frandsen} {et~al.}(2002){Frandsen}, {Carrier}, {Aerts}, {Stello},
  {Maas}, {Burnet}, {Bruntt}, {Teixeira}, {de Medeiros}, {Bouchy}, {Kjeldsen},
  {Pijpers}, \& {Christensen-Dalsgaard}}]{fra02}
{Frandsen}, S., {Carrier}, F., {Aerts}, C., {et~al.} 2002, \aap, 394, L5

\bibitem[{{Grevesse} \& {Noels}(1993)}]{gre93}
{Grevesse}, N. \& {Noels}, A. 1993, in Origin and evolution of the elements:
  proceedings of a symposium in honour of H. Reeves, held in Paris, June 22-25,
  1992. Edited by N. Prantzos, E. Vangioni-Flam and M. Casse. Published by
  Cambridge University Press, Cambridge, England, 1993, p.14, ed.
  N.~{Prantzos}, E.~{Vangioni-Flam}, \& M.~{Casse}, 14

\bibitem[{{Maeder} \& {Meynet}(1989)}]{mae89}
{Maeder}, A. \& {Meynet}, G. 1989, \aap, 210, 155

\bibitem[{{Maeder} \& {Meynet}(2000)}]{mae00b}
{Maeder}, A. \& {Meynet}, G. 2000, \araa, 38, 143

\bibitem[{{Maeder} \& {Meynet}(2001)}]{mae01}
{Maeder}, A. \& {Meynet}, G. 2001, \aap, 373, 555

\bibitem[{{Maeder} \& {Zahn}(1998)}]{mae98}
{Maeder}, A. \& {Zahn}, J.-P. 1998, \aap, 334, 1000

\bibitem[{{Massarotti} {et~al.}(2008){Massarotti}, {Latham}, {Stefanik}, \&
  {Fogel}}]{mas08}
{Massarotti}, A., {Latham}, D.~W., {Stefanik}, R.~P., \& {Fogel}, J. 2008, \aj,
  135, 209

\bibitem[{{Osaki}(1975)}]{osa75}
{Osaki}, J. 1975, \pasj, 27, 237

\bibitem[{{Palacios} \& {Brun}(2007)}]{pal07}
{Palacios}, A. \& {Brun}, A.~S. 2007, Astronomische Nachrichten, 328, 1114

\bibitem[{{Palacios} {et~al.}(2006){Palacios}, {Charbonnel}, {Talon}, \&
  {Siess}}]{pal06}
{Palacios}, A., {Charbonnel}, C., {Talon}, S., \& {Siess}, L. 2006, \aap, 453,
  261

\bibitem[{{Roxburgh}(2005)}]{rox05}
{Roxburgh}, I.~W. 2005, \aap, 434, 665

\bibitem[{{Roxburgh} \& {Vorontsov}(2003)}]{rox03}
{Roxburgh}, I.~W. \& {Vorontsov}, S.~V. 2003, \aap, 411, 215

\bibitem[{{Sweigart} \& {Mengel}(1979)}]{swe79}
{Sweigart}, A.~V. \& {Mengel}, J.~G. 1979, \apj, 229, 624

\bibitem[{{Vogt}(1926)}]{vog26}
{Vogt}, H. 1926, Astronomische Nachrichten, 227, 325

\bibitem[{{Zahn}(1992)}]{zah92}
{Zahn}, J.-P. 1992, \aap, 265, 115

\end{thebibliography}

\end{document}